\documentclass[twocolumn,showpacs,preprintnumbers,amsmath,amssymb]{revtex4}
\newcommand{\un}{~\mathrm}
\newcommand{\ie}{{\it i.e. }}
\newcommand{\eg}{{\it e.g. }}
\newcommand{\unm}{~\mu\mathrm{m}}

\usepackage{graphicx}
\usepackage{dcolumn}
\usepackage{bm}
\usepackage{ulem}
\begin{document}

\preprint{APS/123-QED}

\title{Cleaved surface of {\it i}-AlPdMn quasicrystals:\\Influence of the local temperature elevation at the crack tip on the fracture surface roughness}

\author{L. Ponson}
 \email{ponson@drecam.cea.fr}
\author{D. Bonamy}
\author{L. Barbier}
\affiliation{Fracture Group, Service de Physique et Chimie des Surfaces et Interfaces, DSM/DRECAM/SPCSI, CEA Saclay,
F-91191 Gif sur Yvette, France}

\date{\today}

\begin{abstract}
Roughness of {\it i}-AlPdMn cleaved surfaces are presently analysed. From the atomic scale to 2-3$~\mathrm{nm}$, they are shown to exhibit scaling properties hiding the cluster ($0.45~\mathrm{nm}$) aperiodic structure. These properties are quantitatively similar to those observed on various disordered materials, albeit on other ranges of length scales. These properties are interpreted as the signature of damage mechanisms occurring within a 2-3$~\mathrm{nm}$ wide zone at the crack tip. The size of this process zone finds its origin in the local temperature elevation at the crack tip. For the very first time, this effect is reported to be responsible for a transition from a perfectly brittle behavior to a nanoductile one.
\end{abstract}


\maketitle
\section{Introduction}
Breaking a material to gain information on its microstructure is a common experiment in condensed matter physics. It has been used \eg to measure the surface energy of mica \cite{Obreimoff}, or to determine the weak plans in Silicon monocrystal \cite{Haneman}, and estimate their surface energy \cite{Hauch}. As for fracture surfaces, scientists have also studied their morphology to improve their knowledge of the complex damage and fracture processes occurring at the microstructure scale during the failure of heterogeneous materials. In particular, their roughness was quantitatively analysed with statistical methods and found to be self-affine, characterized by a roughness exponent $\zeta \simeq 0.8$ \cite{Bouchaud4}. Very recent studies have shown that the complete 2D correlation function takes a generic Family-Viseck scaling form characterized by a series of three critical exponents independent of the failure mode and of the nature of the material \cite{Ponson4,Ponson5}. Sandstone \cite{Boffa} and glassy ceramics \cite{Ponson6} were observed to exhibit the same Family-Viseck scaling, but with other critical exponents (in particular $\zeta \simeq 0.4$). The range of length-scales over which the $\zeta\simeq 0.8$ (resp. $\zeta\simeq 0.4$) scaling regime is observed was recently shown to be below (resp. above) the size of the process zone \cite{Bonamy2}.

Thus, the present statements summarize the competition between the effects on fracture surface roughness of the fracture process and of the microstructure: (i) fracture surfaces of {\it disordered} materials are rough with an height correlation function exhibiting universal properties dictated by the universal fracture mechanisms, (ii) fracture surfaces of {\it ordered} materials can be flat under some conditions and the microstructure dominates their cleavage properties. However, wavy and rough cracks were observed in single-crystal silicons as reported in \cite{Deegan}.

To improve such ideas and to better assess the universal properties of fracture surfaces, it is of interest to investigate materials with very different structures and mechanical properties. In return, the very last improvements in the knowledge of the fracture process would allow a better knowledge of material properties. It is with this aim that we turn our interest to revisit the cleavage surface of quasicrystals (QC). Images by Scanning Tunneling Microscopy (STM) of cleaved {\it i}-AlPdMn samples in ultra high vacuum (UHV) were obtained, for cleavage along the 2-fold and 5-fold directions \cite{Ebert, Ebert3} as for decagonal QC \cite{Ebert2}. The apparent roughness of these surfaces was at first interpreted as reminiscent of the bulk cluster structure \cite{Boissieu}. Indeed, it has been widely suggested that the aperiodic structure of QCs with icosahedral symmetry is stabilized by the distribution of stable (pseudo-Bergman or -MacKay) clusters. Ebert et al. \cite{Ebert, Ebert3} suggested that this cluster structure would define a heterogeneity scale revealed by their STM images. The crack was conjectured to propagate along the zones of lower strength in between the so-called clusters, which results in the observed rough fracture surfaces. In contrast, observations by UHV-STM of 5-fold surfaces of {\it i}-AlPdMn at thermal equilibrium (for {\it i}-AlCuFe, see: \cite{Cai}) shows that wide ($> 100 \un nm$) flat terraces with very low corrugation of $40 \un pm$  are present \cite{Barbier} (5-fold first order Fourier amplitude of $2-3 \un pm$). Aperiodic QCs can be seen as the irrational cut of n-dimensional periodic structures \cite{Gratias} ($n=6$ for the icosaedral symmetry). Interpretation of high resolution STM images within these 6-dimensional models allowed identification of the terrace structures. Clearly, the terraces cut the Bergman clusters, revealing an apparent contradiction with the interpretation suggested for cleavage surfaces.

\begin{figure*}[!ht]
\includegraphics[width=0.63\columnwidth]{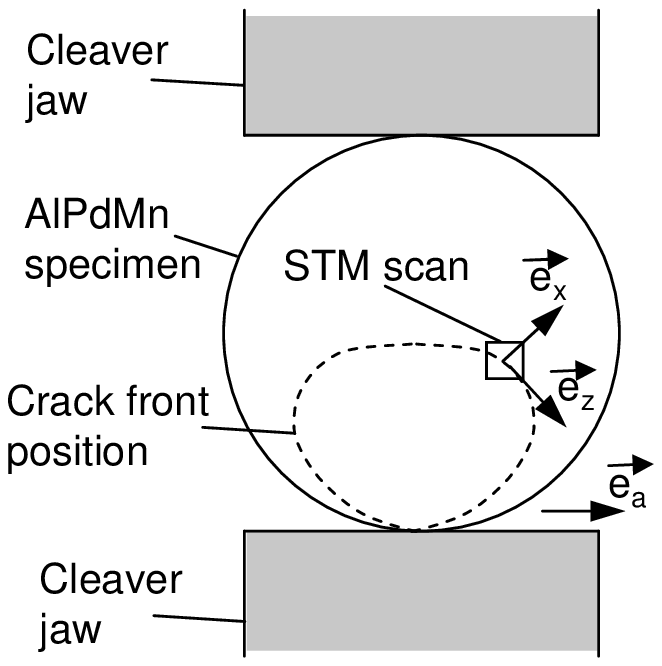}
\includegraphics[width=0.9\columnwidth]{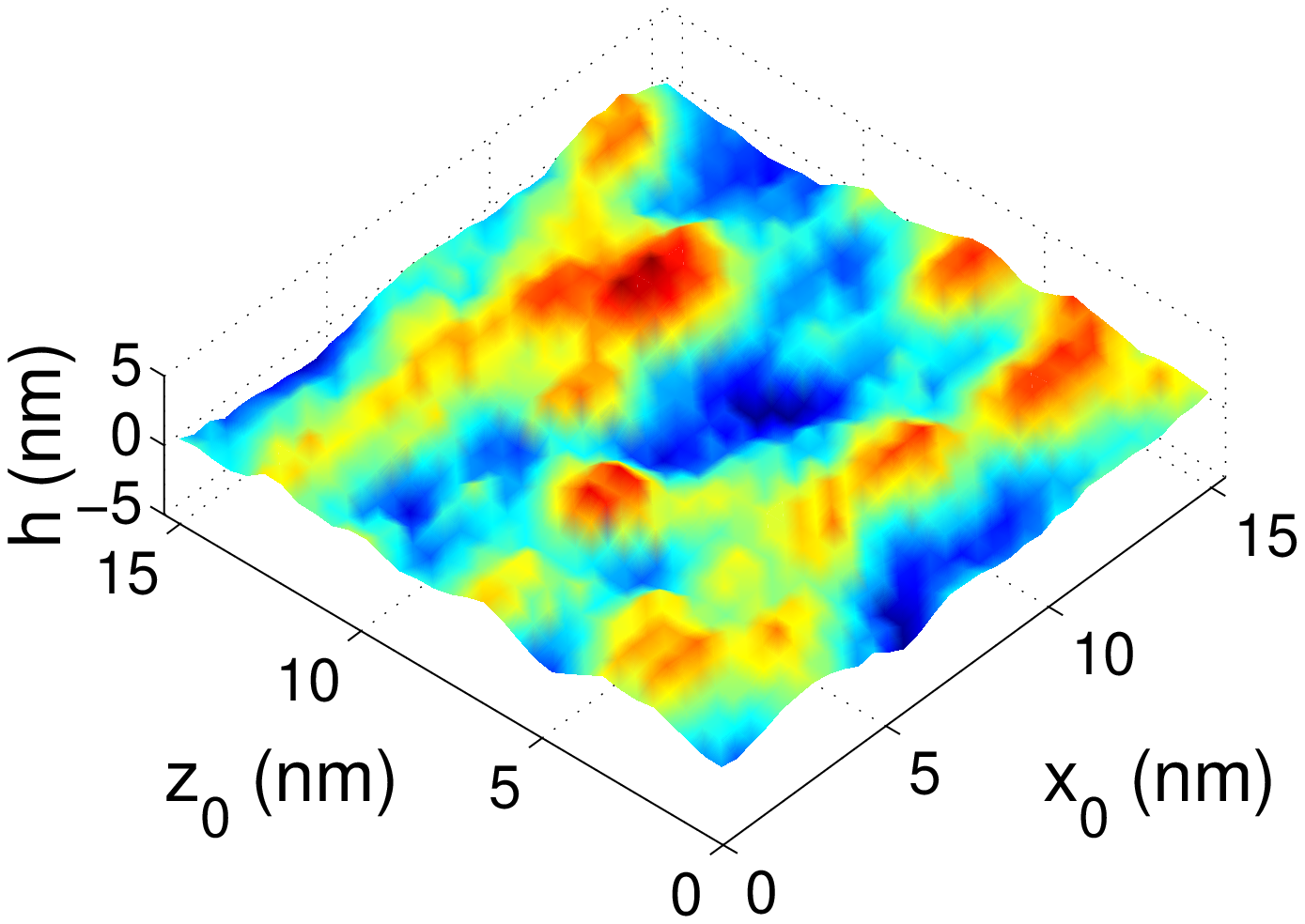}
\centering
\caption{Sketch of the cleavage experiment of an {\it i}-AlPdMn QC (left) and typical topographic image of the resulting fracture surface as observed using STM (right).}
\label{fig1}
\end{figure*}
         
In this paper, we analyse quantitatively the morphology of cleaved {\it i}-AlPdMn QC surfaces at room temperature as measured in 1998 by P. Ebert and co-workers \cite{Ebert, Ebert3}. Following the method recently developped to analyse fracture surfaces of materials \cite{Ponson4,Ponson5}, statistical scaling properties of the roughness of these surfaces can be examined, which is the main point of the present paper. At first, fracture surfaces of these quasi-ordered materials were expected to be reminiscent of the peculiar structure of QCs \cite{Ebert, Ebert3}. Surprisingly, their roughness is shown to follow scaling invariance properties similar to those observed on various disordered materials, irrespective to the cleavage plane. Scale invariance is limited to the range between the atomic scale up to $2-3$ nanometers. Within this range, the structure function is shown to follow {\it quantitatively} the very same universal shape already measured on aluminum alloy, glass and mortar fracture surfaces although this latter involves different length scales. Similarly to conclusions of a previous study \cite{Bonamy2}, we conjecture that the pointed out scaling properties are the signature of non-linear damage processes occurring at the crack tip vicinity during the failure of the {\it i}-AlPdMn QC. In the following, we propose a scenario based on previous calculations \cite{Rice2}, where the local temperature elevation accompanying these non-linear processes generates the nanoplasticity.

\section{Experimental setup}
QC cleavage and fracture surface scanning were performed by P. Ebert and co-workers. Samples of single QC of $Al_{70.5}Pd_{21}Mn_{8.5}$ were cleaved perpendicular to 2-fold and 5-fold axes in ultrahigh vacuum  \cite{Ebert}. To measure the topography of the fracture surfaces, the samples were transferred to a tunneling microscope without breaking the vacuum. 
We estimate lateral resolutions of the STM images to be of the order of $0.05 \un{nm}$ and $0.1 \un{nm}$ parallelly and perpendicularly to the scanning direction respectively \cite{note1}.
Six images ($512 \times 512$ pixels) corresponding to two different cleavage planes were analyzed. Their corresponding scan size and cleavage plane are listed in Table \ref{tab}. Figure \ref{fig1} shows a typical snapshot of a fracture surface of the {\it i}-AlPdMn QC cleaved perpendicularly to the twofold axe. The reference frame of the image $(\vec{e}_{x0},\vec{e}_{h},\vec{e}_{z0})$ is chosen so that $\vec{e}_{x0}$ and $\vec{e}_{z0}$ are respectively perpendicular and parallel to the scanning direction of the STM tip. The scanning direction was chosen in order to scan along the local minimum apparent slope of the surface to avoid rapid vertical motion of the tip. In-plane (along $z_0$ and $x_0$) and out-of-plane (along $h$) length scales of the largest observed features are found to be respectively of the order of 2-3$\un{nm}$ and 1$\un{nm}$. As mentioned by Ebert and coworkers \cite{Ebert}, one can observe peaks and valleys of various size on these fracture surfaces, which suggests a 'fractal' structure.

\section{Experimental results}

\subsection{Self-affine behavior of {\it i}-AlPdMn cleaved surfaces}

\begin{figure}[!ht]
\includegraphics[width=0.9\columnwidth]{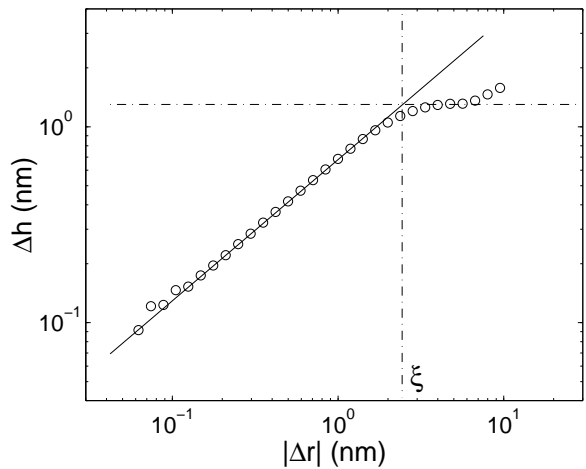}
\centering
\caption{Height-height correlation function $\delta h$ as a function of the distance $|\Delta r|$. The axes are logarithmic. The plain straight line is a power law fit using Eq.~(\ref{eq:self-affine}) with $H=0.72$ and $\ell=0.26\un{nm}$. The self-affine scaling extends up to a cut-off length $\xi \simeq 2.3 \un{nm}$.}
\label{fig2}
\end{figure}

To reveal the scaling properties of QCs surfaces, we computed the height-height correlation function $\Delta h(\Delta \vec{r})$ defined by:

\begin{equation}
\Delta h(\Delta \vec{r})=<(h(\vec{r}_0+\Delta r \;\vec{e}_r)-h(\vec{r}_0))^2>_{\vec{r}_0}^{1/2}
\label{eq:corr}
\end{equation}

\noindent where angular brackets denotes average over all $\vec{r}_0$. To avoid any effect of the inclination of each image on the calculation of the correlation function, its mean plane is substracted before analysis. Figure \ref{fig2} represents variations of $\Delta h$ average over all the directions $\vec{e}_r$ with respect to $\Delta r$ in logarithmic scales for the 2-fold cleaved surface shown in Fig. \ref{fig1}. This curve reveals a clear self-affine behaviour from the atomic scale to a cut-off length $\xi \simeq 2.3\un{nm}$. In other words, for $\Delta r \leq \xi$, $\Delta h$ is found to scale as:

\begin{equation}
\frac{\Delta h}{\ell}=\left(\frac{\Delta r}{\ell}\right)^H
\label{eq:self-affine}
\end{equation}

\noindent where $H \simeq 0.72$ denotes the Hurst exponent and $\ell \simeq 0.26\un{nm}$ refers to the topothesy that is the scale at which $\Delta h$ is equal to $\Delta r$. This scaling property is very robust and measured on all images, irrespective to the scan size and resolution. The value of the Hurst exponent measured here is fairly consistent with the "universal" value $H\simeq 0.8$ widely reported for a broad range of heterogeneous materials \cite{Bouchaud4}. The range of the self-affine behavior includes the radius $\simeq 0.45 \un{nm}$ of the Bergman or Mackay cluster. This means that the morphology of cleaved surfaces in {\it i}-AlPdMn does not reflect the cluster distribution as previously suggested \cite{Ebert}. It reflects more likely some generic failure mechanisms occurring at the microstructure scale and independent on the precise nature of the considered material.

\subsection{Determination of the direction of crack propagation/crack front direction}

To carry on the comparison between the scaling properties of cleaved surfaces in QCs and those observed in other heterogeneous materials \cite{Bouchaud4}, we investigate now their two-dimensional (2D) scaling properties $\Delta h(\Delta \vec{r})$. As a matter of fact, it has been recently shown \cite{Ponson4,Ponson5} that a complete description of the scaling properties of fracture surfaces in heterogeneous materials like \eg glass, metallic alloy, wood or mortar calls for the use of the 2D height-height correlation function. In particular, the direction of the crack propagation $\vec{e}_x$ and the direction of the crack front $\vec{e}_z$ were shown to be the only ones that provides pure scaling properties characterized by two well-defined exponents referred to as $\beta$ and $\zeta$ respectively, while the other directions $\theta$ in between exhibit a combination of the two scaling behaviors (often fitted in coarse approximation by a single power-law characterized by an effective exponent $H(\theta)$ between $\beta$ and $\zeta$).

\begin{figure}[!ht]
\includegraphics[width=0.9\columnwidth]{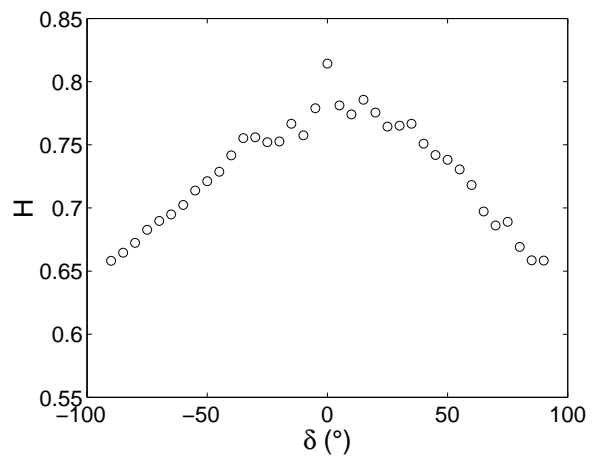}
\centering
\caption{Variation of the effective Hurst exponent $H$ measured along a direction making an angle $\theta$ with the STM scan direction $\vec{e}_{x_0}$. The minimum and maximum (here close to $0^\circ$ and $90^\circ$ respectively) of $H$ co\"{\i}ncide with the direction of crack propagation and the direction of the crack front respectively.}
\label{fig3}
\end{figure}

In a cleavage experiment, like the one giving the present surfaces of interest, the crack is expected propagating within a well-defined plane, but along an {\it a priori} unknown direction (Fig. \ref{fig1}). However, the intrinsic scaling anisotropy of fracture surfaces suggests a method to determine the direction of the crack propagation \cite{Ponson7}. According to this technique, one has to compute the one dimensional (1D) height-height correlation function $\Delta h(\Delta r, \theta) = <(h(\vec{r}_0+\Delta r\vec{e}_r)-h(\vec{r}_0))^2>^{1/2}_{\vec{r}_0}$ along 1D profiles taken along a direction $\vec{e}_r$ making an angle $\theta$ with respect to $\vec{e}_{x_0}$, $\theta=\cos^{-1}(\vec{e_r}.\vec{e}_{x_0})$. The effective exponent $H(\theta)$ defined by $\Delta h(\Delta r, \theta) \propto \Delta r ^{H(\theta)}$ is then estimated (Figure \ref{fig3}). The angle $\theta$ where $H$ is minimum (resp. maximum) is then expected to co\"{\i}ncide with the local direction of the crack propagation $\vec{e}_x$ (resp. the crack front direction $\vec{e}_z$).

Knowing the scanning angle with respect to the laboratory frame $\widehat{(\vec{e}_{cl}, \vec{e}_{x0})}$ , the resulting angle $\widehat{(\vec{e}_{cl},\vec{e}_x)}$ between the direction $\vec{e}_{cl}$ parallel to the side of the cleaver and the crack propagation direction $\vec{e}_x$ can be estimated using this procedure and is listed in Table \ref{tab}.
For all the images, except image $\sharp$ 5 \cite{noteImage5}, the crack was found to have propagated within an angle of $35^\circ \pm 5^\circ$ with respects to the sides of the cleavers. This direction was found to be uncorrelated with the tip scanning direction. In other words, the anisotropy measured on the fracture surface is well a signature of the crack propagation direction and is not induced by the imaging technique.

\subsection{2D Family-Viscek scaling of the {\it i}-AlPdMn cleaved surfaces}

Once the direction of crack propagation $\vec{e}_x$ and the crack front direction $\vec{e}_z$ clearly identified in the various STM images, the 2D scaling properties of the fracture surfaces can be properly characterized.

\begin{figure}[!h]
\includegraphics[width=0.9\columnwidth]{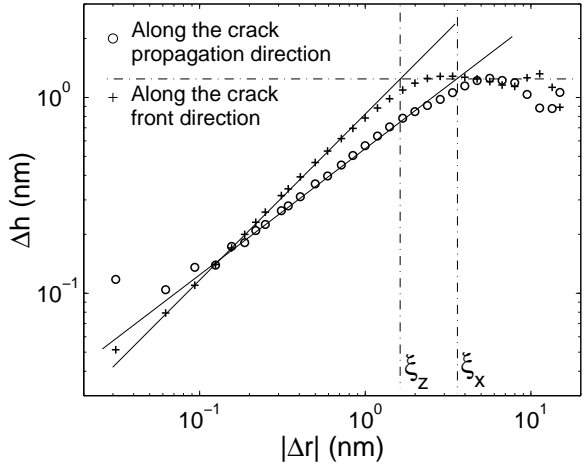}
\centering
\caption{1D height-height correlation function measured parallel to the expected crack propagation direction and the crack front of the QC fracture surface presented Fig. \ref{fig1}. The axes are logarithmic. The plain straight lines are power law fits using Eq.~(\ref{eq:1Dcorr}) with $\zeta \simeq 0.83$, $\beta \simeq 0.65$, , $\ell_z \simeq 0.29\un{nm}$ and $\ell_x \simeq 0.19\un{nm}$.}
\label{fig4}
\end{figure}

First, the 1D height-height correlation functions $\Delta h(\Delta x)=<(h(z,x+\Delta x)-h(z,x))^2>^{1/2}$ along the direction of the crack propagation $\vec{e}_x$, and $\Delta h(\Delta z)=<(h(z+\Delta z,x)-h(z,x))^2>^{1/2}$ along the crack front direction $\vec{e}_z$ are computed. They are represented in logarithmic scales in Fig. \ref{fig4} for the 2-fold cleaved surface shown in Fig. \ref{fig1}. The two correlation functions follow a power-law behavior according to: 

\begin{equation}
\frac{\Delta h(\Delta x)}{\ell_x} = \left(\frac{\Delta x}{\ell_x}\right)^\beta, \quad
\frac{\Delta h(\Delta z)}{\ell_z} = \left(\frac{\Delta z}{\ell_z}\right)^\zeta
\label{eq:1Dcorr}
\end{equation}

\noindent where the two scaling exponents are $\zeta \simeq 0.83$ and $\beta \simeq 0.65$. These power laws are followed for a range between atomic distances up to $\xi_z \simeq 1.6\un{nm}$ and $\xi_x \simeq 3.2\un{nm}$ defining cut-off lengths along the $z$-axis and the $x$-axis respectively. The measured values of $\zeta$, $\beta$, $\xi_z$ and $\xi_x$ on the six analyzed STM images are listed in Table \ref{tab}. The scaling exponents measured on the two family of QC fracture surfaces are found to be independent of the cleavage direction (5-fold and 2-fold) and their value are very close to $\zeta \simeq 0.8$ and $\beta \simeq 0.6$ already measured on other disordered materials \cite{Ponson5, Ponson4}. The cut-off lengths are found to be $\xi_z = 2\un{nm} \pm 0.5\un{nm}$ and $\xi_x = 3\un{nm} \pm 0.5\un{nm}$ irrespective to the scan size and the cleavage direction.

\begin{table*}
\caption{\label{tab}Scan size, cleavage plane, measured angle $\widehat{(\vec{e}_{cl}, \vec{e}_{x})}$ between the direction $\vec{e}_x$ of crack propagation and $\vec{e}_{cl}$ unit vector along the cleaver side. $\zeta_{1D}$ and $\beta_{1D}$: Scaling exponents measured from the calculation of the 1D correlation function. $\zeta_{2D}$, $\beta_{2D}$ and $z$: Scaling exponents for the 2D correlation function. $\xi_z$ and $\xi_x$: Cut-off lengths along the $z$ and $x$-axis respectively. Error bars are the standard deviation.}
\begin{ruledtabular}
\begin{tabular}{cccccccccccc}
 Image & Scan size & Cleavage & $\widehat{(\vec{e}_{cl}, \vec{e}_{x})}$ & $\zeta_{1D}$ & $\zeta_{2D}$ & 
$\beta_{1D}$ & $\beta_{2D}$ & $z$ & $\xi_z$ & $\xi_x$\\ 
 &($\un{nm}^2$) & plane & & & & & & & (nm) & (nm) \\
\hline
 $\sharp$ 1 & 15 $\times$ 15 & 2-fold & $30^\circ$ & 0.83 & 0.76 & 0.65 & 0.67 & 1.14 & 1.6 & 3.2 \\
 $\sharp$ 2 & 7.5 $\times$ 7.5 & 2-fold & $30^\circ$ & 0.84 & 0.74 & 0.64 & 0.61 & 1.24 & 1.9 & 3.2 \\
 $\sharp$ 3 & 75 $\times$ 75 & 2-fold & $30^\circ$ & 0.78 & 0.73 & 0.71 & 0.65 & 1.13 & 2.5 & 2.8 \\
 $\sharp$ 4 & 30 $\times$ 30 & 5-fold & $35^\circ$ & 0.79 & 0.79 & 0.70 & 0.69 & 1.16 & 1.7 & 2.8 \\
 $\sharp$ 5 & 75 $\times$ 75 & 5-fold & $-10^\circ$ & 0.79 & 0.71 & 0.75 & 0.68 & 1.12 & 2.6 & 3.7 \\
 $\sharp$ 6 & 75 $\times$ 75 & 5-fold & $40^\circ$ & 0.82 & 0.74 & 0.73 & 0.61 & 1.28 & 2.1 & 3.2 \\
\hline
 Average\footnote{Without Image $\sharp$ 5} & & & $35^\circ \pm 5^\circ$ & 0.81 $\pm$ 0.03 & 0.76 $\pm$ 0.03 & 0.68 $\pm$ 0.05 & 0.65 $\pm$ 0.04 & 1.20 $\pm$ 0.07 & 2.1 $\pm$ 0.5 & 3.2 $\pm$ 0.5 \\
\hline
\hline
Final \footnote{2D analysis}& \multicolumn{2}{c}{$\zeta \simeq$ 0.76}&\multicolumn{2}{c}{$\beta \simeq$ 0.65}&\multicolumn{2}{c}{$z \simeq$ 1.20}&\multicolumn{2}{c}{$\xi_z \simeq 2.1\un{nm}$}&\multicolumn{2}{c}{$\xi_x \simeq 3.2\un{nm}$}\\
\end{tabular}
\end{ruledtabular}
\end{table*}

\begin{figure}[!h]
\includegraphics[width=0.9\columnwidth]{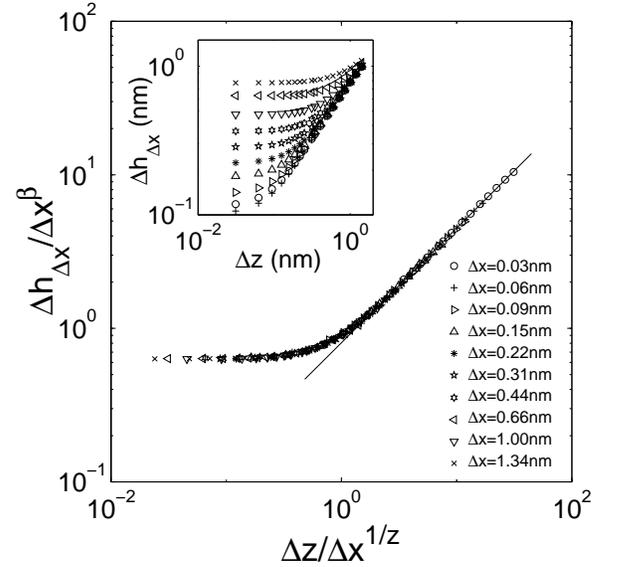}
\centering
\caption{Inset: 2D height-height correlation function $\Delta h_{\Delta x}(\Delta z)$ for various angles (i.e. values of $\Delta x$/$\Delta z$). Using the indicated scaling properties with the exponents reported in Table \ref{tab}., a collapse of the data on a single curve is readily obtained}
\label{fig5}
\end{figure}

Second, the complete 2D height-height correlation function defined by Eq.~(\ref{eq:corr}) expressed in the frame ($\vec{e}_x$,$\vec{e}_z$) is investigated. The correlation function $\Delta h_{\Delta x}$ of the image presented in Fig. \ref{fig1} are plotted as a function of $\Delta z$ in the inset of Fig. \ref{fig3}. To recover the 1D scaling given by Eq.~(\ref{eq:1Dcorr}) when $\Delta x=0$ and $\Delta z=0$, $\Delta h$ and $\Delta z$ are normalized by $\Delta x^\beta$ and $\Delta x^{1/z}$ with $z=\zeta/\beta$ in the main graph of Fig. \ref{fig5}. All the curves are found to collapse on a single master curve which is characterized by a plateau regime followed by a power law variation with exponent $\zeta=0.75$. In other words:

\begin{equation}
\begin{array} {l}
   \Delta h(\Delta z,\Delta x)\propto \Delta x^{\beta}f(\Delta z/\Delta x^{1/z}) \\
\\
$where$\quad	f(u) \propto \left\{
\begin{array}{l l}
1 & $if u$ \ll c  \\
u^{\zeta} & $if u$ \gg c
\end{array}
\right.
\end{array}
\label{eq:2dcorr}
\end{equation}

\noindent for length-scales $\Delta x \leq \xi_x$ and $\Delta z \leq \xi_z$. This scaling, referenced to as Family-Viscek scaling \cite{Family}, corresponds precisely to the one exhibited - at different length-scales but with the similar scaling exponents $\zeta$, $\beta$ and $z=\zeta/\beta$ - by heterogeneous materials \cite{Ponson4, Ponson5}.

Finally, one can quantitatively compare the 2D scaling of the fracture surfaces of {\it i}-AlPdMn within the  $~ 0.1\un{nm}$ to $~ 3\un{nm}$ range to the 2D scaling obtained for glass, aluminum alloy and mortar, even the range of distances is very different (respectively, from $~ 1\un{nm}$ to $~ 100\un{nm}$, $~ 1~\mu\mathrm{m}$ to $~ 100~\mu\mathrm{m}$ and $~ 10~\mu\mathrm{m}$ to $~ 1~\un{mm}$). In order to get rid of the very different length scales for the different materials, Eq.~(\ref{eq:2dcorr}) is rewritten as a function of the non-dimensionless quantities $\Delta x/\ell_x$ and $\Delta z/\ell_z$. This leads to:

\begin{equation}
\begin{array} {l}
\Delta h(\Delta z,\Delta x) =\ell_x (\Delta x/\ell_x )^{\beta} g \left( u= \frac{\ell_z}{\ell_x}\frac{\Delta z/\ell_z}{(\Delta x/\ell_x)^{1/z}} \right)\\
\\
$where$ \quad g(u) =
\left\{
\begin{array}{l l}
1 & $if u$ \ll 1  \\
u^{\zeta} & $if u$ \gg 1
\end{array}
\right.
\end{array}
\label{eq:2dcorrnd}
\end{equation}

\begin{figure}[!ht]
\includegraphics[width=0.9\columnwidth]{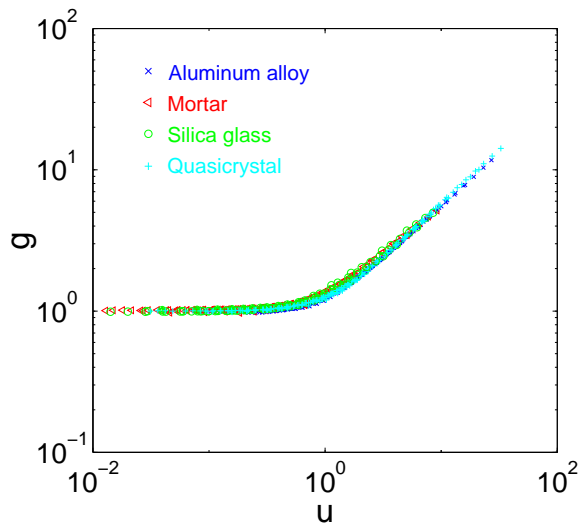}
\centering
\caption{Scaling function $g(u)$ involved in Eq.~(\ref{eq:2dcorrnd}) and measured on four different materials, glass, metallic alloy, mortar and QC.}
\label{fig6}
\end{figure}

\noindent The master curves $g(u)$ obtained in previous studies \cite{Ponson5,Ponson6} for aluminum alloy, mortar and silica glass fracture surfaces and the one measured here for the {\it i}-AlPdMn QC co\"{\i}ncide as shown on Fig. \ref{fig6}. The scaling function is found to be independent of the material, not only in the plateau regime and the power law variation as predicted by the values of the measured exponents, but also within the crossover domain.

\section{Discussion}

The present analysis clearly shows that the scaling properties exhibited by cleaved surface of {\it i}-AlPdMn shares many properties with the ones of other disordered materials:

\begin{itemize}
\item[(i)] The cleaved surface are self-affine characterized by an effective Hurst exponent close to $0.8$ as reported for a wide range of heterogeneous materials \cite{Bouchaud4}.
\item[(ii)] The 2D correlation function obeys a Family-Viscek scaling [Eq.~(\ref{eq:2dcorr})] characterized by three critical exponent $\zeta \simeq 0.8$, $\beta \simeq 0.65$ and $z=\zeta/\beta$ as previously observed in glass, metallic alloy, wood and mortar \cite{Ponson5,Ponson4}.
\item[(iii)] Following the Family-Viscek scaling laws, one unique crossover function $g(u)$ is obtained for all materials.
\end{itemize}

This clearly questions the interpretation proposed in \cite{Ebert} where the morphology of cleaved surfaces in icosahedral QCs was interpreted as the signature of the cluster based stucture of QCs; the crack front having propagated along the weaker zones in between. Origin of the morphology of the fracture surface would be found more likely in the failure mechanisms, common with other heterogeneous materials albeit occurring at different scales. 

The origin of the scaling properties of fracture surfaces in heterogeneous materials and the universality of the measured scaling exponents is still matter of debate. Hansen and Schmittbuhl \cite{Hansen} suggested that the universal scaling properties of fracture surfaces result from the propagation of the fracture front following a damage coalescence process described by a stress-weighted percolation phenomenon in a self-generated quadratic damage gradient. This approach succeeds to reproduce self-affine fracture surfaces, but fails to capture the Family-Viscek scaling given by Eq.~(\ref{eq:2dcorr}). Bouchaud et al. \cite{JPBouchaud} proposed to model the fracture surface as the trace left by a crack front moving through randomly distributed microstructural obstacles - the dynamics of which is described through a phenomenological nonlinear Langevin equation, keeping only the terms allowed by the symmetry of the system. This approach capture the Family-Viseck scaling given by Eq.~(\ref{eq:2dcorr}), but predicts the existence of two self-affine regimes with a crossover length that diverges as the crack growth velocity vanishes \cite{Daguier}, in contradiction with experimental observations \cite{Bonamy2}. Ramanathan et al. used Linear Elastic Fracture Mechanics (LEFM) to derive a linear non-local Langevin equation within both elasto-static \cite{Ramanathan} and elasto-dynamic \cite{Ramanathan2} approximations which succeeds to reproduce scale invariant crack surface roughness in qualitative - but unfortunately not quantitative - agreement with the experimental observations. Recently, Bonamy et al. \cite{Bonamy2} showed that damage screening at the crack tip may explain the critical exponents measured on the fracture surfaces of heterogeneous materials. In this approach, the size of the process zone $R_c$ is expected to be the relevant length scale setting the upper cut-off length $\xi$ that limits the Family-Viseck scaling as given by Eq.~(\ref{eq:2dcorr}) with $\zeta \simeq 0.75$, $\beta\simeq 0.6$ and $z=\zeta/\beta\simeq 1.2$. This prediction was confirmed experimentally for Silica glass \cite{Bonamy2}. For the presently analysed QC cleavage surfaces, the Family-Viseck scaling properties is observed at scales smaller than $\xi_z=2-3\un{nm}$. This length $\xi_z$ can be therefore interpreted as the size of the process zone $R_c$.

On the theoretical side, presence of a process zone in aperiodic structure was indeed observed in molecular dynamics simulations of a 2D model with Lenard-Jones interaction potentials \cite{Mikulla}. This process zone would result from effective dislocation creation and motion like in monocrystals \cite{Brede}. The present measurement of the process zone extend $R_c$ can be compared with a straightforward estimation based on the Dugdale model for a cohesive zone at a crack tip \cite{Dugdale}:
\begin{equation}
Rc \simeq \frac{8}{\pi}(\frac{K_{Ic}}{\sigma_Y})^2
\label{eq7}
\end{equation}
Cleaved surfaces of QCs were obtained at room temperature. Toughness is not a so easy quantity to measure: early values were found from $0.3$ at $300 \un{K}$ up to $3 \un{MPa.m^{1/2}}$ at $600 \un{K}$ \cite {Yokoyama} whereas, comparing the most recent toughness measurements for similar aluminium based {\it i}-QCs, one gets: $\simeq 1.25\un{MPa.m^{1/2}}$, {\it i}-AlPdMn \cite{Deus} or $\simeq 0.89\un{MPa.m^{1/2}}$, {\it i}-AlCuFe \cite{Giacometti}. One takes the reasonable value for $K_{Ic}$ of $1\un{MPa.m^{1/2}}$. The room temperature yield strength can be estimated to $\simeq E/3$ ($E=194\un{GPa}$ \cite{Feuerbacher2} being the Young's modulus) giving a value of $\sigma_Y \simeq 65\un{GPa}$. Thus, one gets $R_c \simeq 0.09\un{nm}$ clearly too small to reproduce the experimental value of $2-3\un{nm}$ observed experimentally on fracture surfaces. Such a very low process zone size of the order of the atomic scale would correspond to a perfectly brittle rupture.

However, room temperature parameters were used in the above calculation, whereas a temperature rise within and around the process zone must be expected. To solve such a discrepancy between the measured and calculated process zone sizes, we propose to invoke the temperature rise at the crack tip during the cleavage test and its effect on the fracture mode of icosahedral QC. Kraft and Irwin \cite{Kraft} predicted that such a temperature effect could change the fracture behavior of some materials. This has been actually observed experimentally in highly ductile materials such as aluminum alloy and steel \cite{Ravichandran} where a temperature rise as high as $200\un{K}$ was measured whithin a $100\unm$ size zone. For QC, the much smaller value of $R_c$ measured on fracture surfaces allows expecting an even higher temperature rise at the vicinity of the crack tip. This could play a crucial role. Rice and Levy \cite{Rice2} proposed calculations to assess the temperature field at the tip of a crack propagating at constant velocity $v_{crack}$. The temperature at the crack tip is given by:
\begin{equation}
T_{tip}(R_c)=T_{room}+0.366\frac{1-\nu^2}{E}\frac{K_{Ic}^2}{\sqrt{\rho c k}}\sqrt{\frac{v_{crack}}{R_c}}
\label{eq3}
\end{equation}
\noindent For the experimental fracture conditions ($T=300\un{K}$), one takes for a {\it i}-AlPdMn single QC: the Poisson's ratio $\nu=0.38$ \cite{Feuerbacher2}, the mass density $\rho=5000\un{kg.m^{-3}}$, the specific heat $c=540\un{J.kg^{-1}.K^{-1}}$\cite{Legault, Walti} and the heat conductivity $k=1.5\un{W.m^{-1}.K^{-1}}$ \cite{Chernikov} that are weakly $T$ dependent above room $T$. The temperature is calculated to be constant until a distance $R_T$ from the crack tip and then expected to decrease as 1/r. $R_T$ is given by \cite{Rice2}:
\begin{equation}
R_T=2\sqrt{\frac{k}{\rho c}}\sqrt{R_c/v_{crack}}
\label{eq4}
\end{equation}
If damage occurs at the crack tip and thus $R_c$ is non-zero, one expects a temperature rise that will enable ductile fracture at the crack tip. Indeed, the yield stress of {\it i}-AlPdMn has been observed to decrease very fastly with the temperature. Computing various experimental results \cite{Feuerbacher} \cite{Brunner},\cite{Takeuchi}, one can derive the phenomenological temperature dependence within $300-1300\un{K}$:
\begin{equation}
\sigma_Y(T)=3.5.10^{11}e^{-0.0067 T} \un{(MPa)}
\label{eq5}
\end{equation}
Finally, the process zone extension will be given by the distance from the crack tip for which the mean stress resulting from the main crack is equal to the yield stress of the QC at some temperature. This leads to:
\begin{equation}
\sigma_Y(T(R_c))=K_{Ic} \sqrt{\frac{\pi}{8 R_c}}
\label{eq6}
\end{equation}
where $\sigma_Y$ depends on the distance $r$ to the tip because of the non-uniform temperature field $T(r)$ at the vicinity of the crack tip. Solving Eq.~(\ref{eq6}) thanks to the dependance of $\sigma_Y$ with $T$ and the temperature field $T_{tip}(r)$ given by Eq.~(\ref{eq5}), (\ref{eq3}) and (\ref{eq4}), one gets one unique solution $R_c \simeq 4\un{nm}$ irrespective to the crack speed in the range $v_{crack}>v_{sound}/10$ where $v_{sound}=4000\un{m.s^{-1}}$ is the sound velocity of the quasicrystal. This new estimation of $R_c$ is now in agreement with the process zone size measured on the quasicrystal fracture surfaces. The temperature at the crack tip is calculated to be of the order of $540\un{K}$ while the zone size of constant temperature is found to be of the order of $R_T=4.8\un{nm}$ [Eq.~(\ref{eq4})]. Thus, $R_T$ is very close to $R_c$ and the temperature is almost uniform over the whole process zone.

Equation (\ref{eq6}) provides the equilibrium position of the system. It must be emphasized that the situation corresponding to a perfectly brittle failure with $R_c \simeq 0.1\un{nm}$, smaller than the atomic lengthscale, would be the equilibrium position of the system if no plastic failure mechanisms (and thus no local elevation temperature) occur during the crack propagation. However, irreversible processes refered to as fracto-emission \cite{Lawn} are systematically involved as previously observed during perfectly brittle failure of mica samples \cite{Obreimoff}. These mechanisms generate a local temperature elevation that makes the uniform temperature field equilibrium position unstable. Upon fracture propagation, this local temperature rise makes the material softer (with a strong reduction of $\sigma_Y(T)$) allowing the process zone to extend up to a few $\un{nms}$ stabilizing in that way the process zone temperature. Further, it is the enhanced $R_c$ value that governs the fracture surface roughness. It is finally worth noting that if the temperature elevation is quite hight for {\it i}-AlPdMn QC, its $R_c$ value is smaller than for heterogeneous materials (glass, mortar... ) for which a moderate temperature increase can be expected.

\section{Conclusion}

As measured from STM images \cite{Ebert}, the roughness of cleaved surface of {\it i}-AlPdMn has been revisited. Contrary to the previous interpretation, no signature of the Bergman or Mackay cluster have been revealed: From atomic scale up to $2-3\un{nm}$, Family-Vicsek scaling properties similar to the ones traditionnally observed on various disordered materials below the size of the process zone are observed. The cut-off length scale is however unusual ($<5\un{nm}$). It is interpreted as the size of the process zone providing that the local heating at the crack tip is well taken into account. These properties, interpreted to be the signature of plastic failure mechanisms, suggest that the failure of {\it i}-AlPdMn QC at $300\un{K}$ is ductile within the $2-3\un{nm}$ from the crack tip. We propose a scenario that predicts a nanometric process zone size where the local temperature elevation accompanying these non-linear processes generates this nanoplasticity. In other words, the temperature dependance of the yield stress explains the transition from a flat (or cluster driven) plane to a rough one. For the very first time, the intrinsic temperature effect of the fracture process is reported to be responsible for a transition from a perfectly brittle behavior to a nanoductile one. 

\begin{acknowledgments}
We thank P. Ebert and co-workers for kindly having providing us their very valuable raw data of STM images of quasicrystal cleavage surfaces and P. Ebert for careful reading of the manuscript. We are also grateful to E. Bouchaud and J. P. Bouchaud for enlightening discussions. 
\end{acknowledgments}


\begin{thebibliography}{40}
\bibitem{note1}
Here, we assume that the lateral resulation of the STM images is given by the length scale at which the 1D correlation function deviates from the power-law behavior at low scales (see Fig. \ref{fig2} and \ref{fig4}).
\bibitem{noteImage5} It is worth to note that the analysis performed on the image $\sharp$ 5 that does not seem to lead to a realistic crack propagation direction involves the largest image making poor the available statistics at the relevant length scales, \ie between 0.2 and 3$\un{nm}$.
\end{thebibliography}
\end{document}